\documentclass[]{jot}

\usepackage{multirow}
\usepackage{booktabs}
\usepackage{graphicx}
\usepackage[utf8]{inputenc}
\usepackage[T1]{fontenc}
\usepackage[english]{babel}
\usepackage{microtype} 
\usepackage{tabularx} 
\usepackage{booktabs} 
\usepackage{pifont}
\usepackage{graphicx}
\usepackage{textcomp}
\usepackage{listings}
\usepackage{xcolor}
\usepackage{epsfig}
\usepackage{wrapfig}
\usepackage{multirow}
\usepackage{lmodern}
\usepackage{placeins}
\usepackage{rotating}
\usepackage{wrapfig}
\usepackage{subcaption}
\jotdetails{
    volume=19,     
    articleno=e1,  
    year=2020,  
    license=ccby   
}

\articletype{manual} 

\title{Fixing Multiple Type Errors in Model Transformations with Alternative Oracles to Test Cases}

\author{Zahra VaraminyBahnemiry}
\author{Jessie Galasso}
\author{Houari Sahraoui}
\affil{University of Montreal, Montreal, Canada}

\runningtitle{Fixing Multiple Type Errors in Model Transformations with Alternative Oracles to Test Cases} 
\runningauthor{Z. VaraminyBahnemiry \textit{et al.}}

\begin{abstract}
	
This paper addresses the issue of correcting type errors in model transformations in realistic scenarios where neither predefined patches nor behavior-safe guards such as test suites are available. 
Instead of using predefined patches targeting isolated errors of specific categories, we propose to explore the space of possible patches by combining basic edit operations for model transformation programs. 
To guide the search, we define two families of objectives: one to limit the number of type errors and the other to preserve the transformation behavior.
To approximate the latter, we study two objectives:

 minimizing the number of changes and keeping the changes local.
Additionally, we define four heuristics to refine candidate patches to increase the likelihood of correcting type errors while preserving the transformation behavior. 
We implemented our approach for the ATL language using the evolutionary algorithm NSGA-II, and performed an evaluation based on three published case studies. The evaluation results show that our approach was able to automatically correct on average more than 82\% of type errors for two cases and more than 56\% for the third case.
\end{abstract}

\begin{document}
\maketitle
\thispagestyle{firststyle}
\marginmark
\section{Introduction}

Model-Driven Engineering (MDE) is increasingly used for product development in industries like automotive, telecom or 
banking~\cite{Whittle2014State}. In those industries, the primary interest in modeling recently shifted from producing 
complex models -- mainly for documenting software systems -- to using these models to 
(semi-)automatically generate software artifacts by means of model transformations~\cite{combemale2016engineering}. 

Model transformations usually take as input models expressed in a modeling language (i.e., metamodel), which can be of general-purpose 
(e.g., UML) or domain-specific (e.g., AUTOSAR\footnote{\url{http://www.autosar.org}}  for automotive systems).
The outputs of model transformations can be either models (possibly conforming to different metamodels), or texts such as source code or XML documents.
In this paper, we focus on the former, i.e., model-to-model transformations.
Model transformation programs can be written in general programming languages or transformation-dedicated languages such as ATL~\cite{jouault2008atl}.
These programs usually describe transformation rules that indicate how to transform elements of the input models into elements
of the output models.

Whether they are learned automatically from examples, like in~\cite{baki2016multi}, or written manually, these transformations must be checked
to ensure they are free of errors.
Transformation languages such as ATL are dynamically typed, making transformations expressed in these languages particularly prone to
type errors, such as referring to elements that do not exist in the metamodels, or initializing properties with values of the wrong types.

A way to automatically correct type errors is to provide predefined patches for each category of errors~\cite{cuadrado2018-quickfix}.
Although this approach may be useful for developers, it suffers from two limitations.
Firstly, predefined patches require an intensive knowledge  to modify them or to define new ones (e.g., for new categories of errors). 
Secondly, they fix errors individually without taking into account possible interactions between them~\cite{cuadrado2018-quickfix} 
and may thus introduce new errors while trying to fix existing ones.
Another way to tackle the correction of type errors in transformations is to use automatic program repair techniques such as search-based algorithms. These techniques have been proven to efficiently
support developers for debugging and correction tasks~\cite{monperrus2018-progrepairbiblio}.
Contrary to predefined patches, they enable to explore a space of potential patches, and may help overcome the aforementioned limitations.
These techniques  closely relate to oracles
 checking whether the program behavior is correct after applying a  patch, test suites being the most popular 
oracles~\cite{monperrus2018-progrepairbiblio}.
However, a substantial amount of knowledge is required to provide representative test suites that would
constitute a relevant oracle~\cite{staats2011-testingfoundationsrevisited}, especially for transformations
that use complex structures as input/output. 
Moreover, in the specific context of type errors, valuable patch solutions may fix most of the errors but not all, and the resulting 
transformation cannot be executed -- and then be tested -- as type errors are syntactic errors.
Relying on test suites to guarantee that type error patches preserve a transformation behavior is thus hardly possible.

In this paper, we define a method for patch recommendation fixing type errors in model transformation programs without  relying on 
predefined patches nor test suites.
This method does not seek fully-automatic correction, but rather to alleviate developers' tasks by avoiding patch maintenance and test suites definition.
Thus, its goal is to recommend to developers patches correcting the most errors possible while preserving the behavior of a given faulty transformation.
In a first phase, we propose to explore the space of possible combinations of basic edit operations to find the sequences (i.e., patches) 
that  repair several type errors simultaneously.
To limit the transformation's behavior deviation, we explore the idea of using several objectives to guide the search,
 as surrogate to test oracles.
We test two objectives we think are behavior-preserving: a) minimizing the changes introduced by the patches and b) preserving the 
transformation footprint with respect to the involved input/output languages. We analysed the behavior of faulty transformation programs corrected by this first phase and identify four types of recurring behavior deviations, along with the edit operations introducing them, which may be prevented by following simple guidelines.
However, implementing these guidelines in objectives would be too resource-consuming and make the method non-tractable.
Thus, we define four heuristics to improve
the decisions made during the exploration phase and apply them once, in a second phase, on the best patches obtained in the 
first phase, to further prevent possible behavior deviations.

We evaluate these two phases using three existing ATL model-to-model transformations, with a published dataset containing several mutations of these transformations with various errors and error categories.
The evaluation of the first phase showed contrasting results: while we succeeded to correctly fix, on average, respectively 80\% and 73\% of the type errors while preserving a correct behavior for two transformations, this correction rate was lower (36\%) for the third transformation. However, after applying the heuristics during the second phase, the correction rates increased, on average, to 83\%, 82\% and 57\%, respectively.

We made the following contributions:
\begin{itemize}
\item We adapt an evolutionary population-based algorithm to automatically generate patches which can fix several type errors at the same time in model transformation programs;
\item We show that two objectives (namely, minimizing the changes and keeping the changes local) help to guide the patch generation to preserve the behavior of a corrected model transformation program;
\item We define four heuristics to refine the obtained patches and show that these heuristics further limit behavior deviation.
\end{itemize}

The remainder of this paper is organized as follows.
Section~\ref{sec:problem} gives the necessary background and discusses issues related to automatically fixing type errors in model transformations.
Section~\ref{sec:approach} describes the two-step approach to fix type errors without predefined patches nor test cases.
An implementation and an evaluation of our approach are provided in Section~\ref{sec:evaluation}.
Section~\ref{sec:rw} presents related work. We discuss our findings and conclude in Section~\ref{sec:conclusion}.

\section{Background}\label{sec:problem}

In this section, we start by giving some background information about ATL and type errors in ATL transformation programs.
Then, we present the challenges of repairing those transformations. Finally, we present NSGA-II~\cite{deb2000-nsga}, the evolutionary population-based algorithm we use in our approach.

\subsection{Type Errors in ATL Transformations}\label{sec:context}

\begin{figure*}[ht!]
    \centering
    	\captionsetup{skip=1pt}
    \begin{subfigure}[t]{0.5\textwidth}
        \centering
        \includegraphics[width=0.99\linewidth]{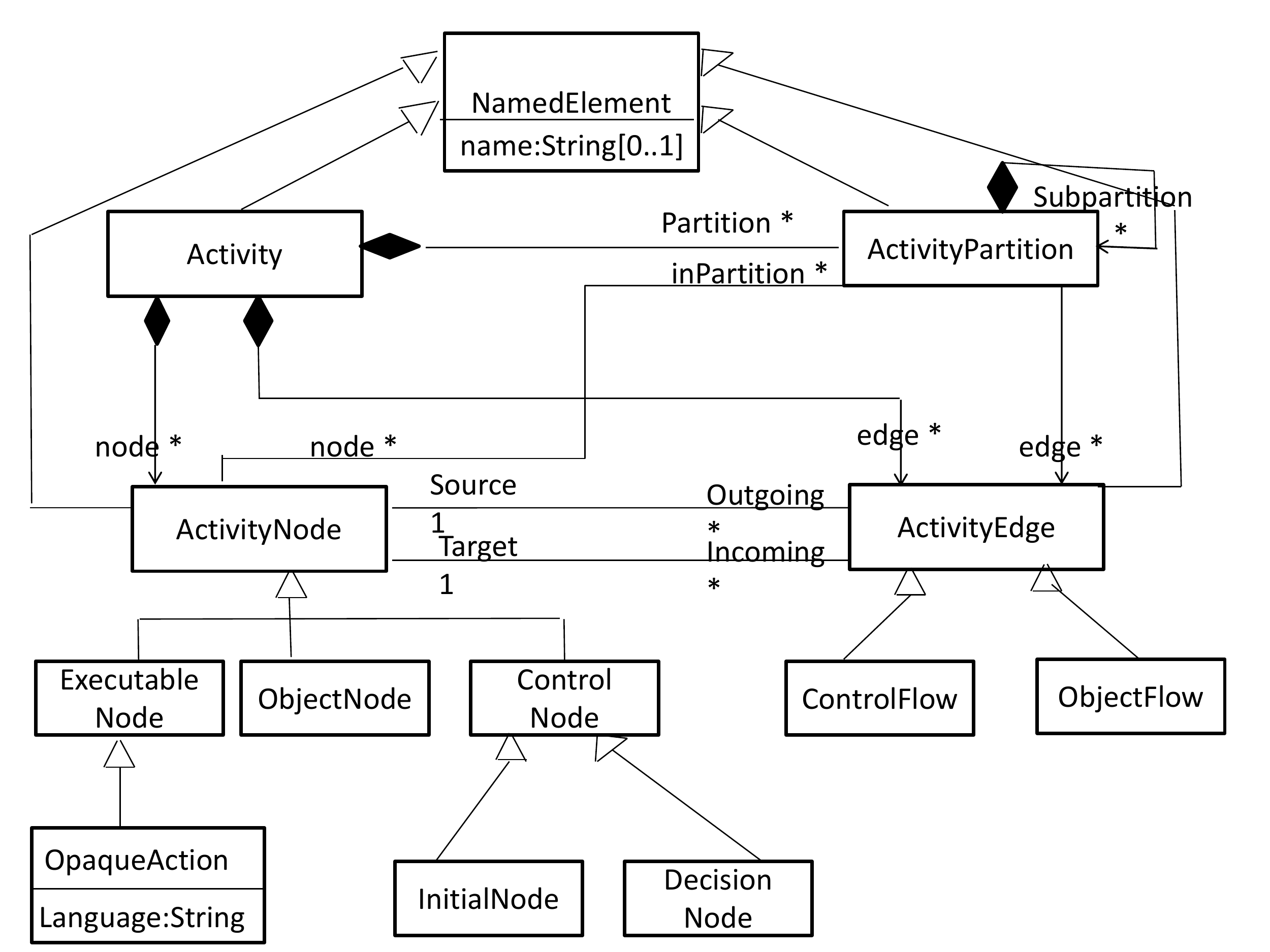}
        \caption{UML AD metamodel}
    \end{subfigure}%
    ~
    \begin{subfigure}[t]{0.5\textwidth}
        \centering
        \includegraphics[width=0.99\linewidth]{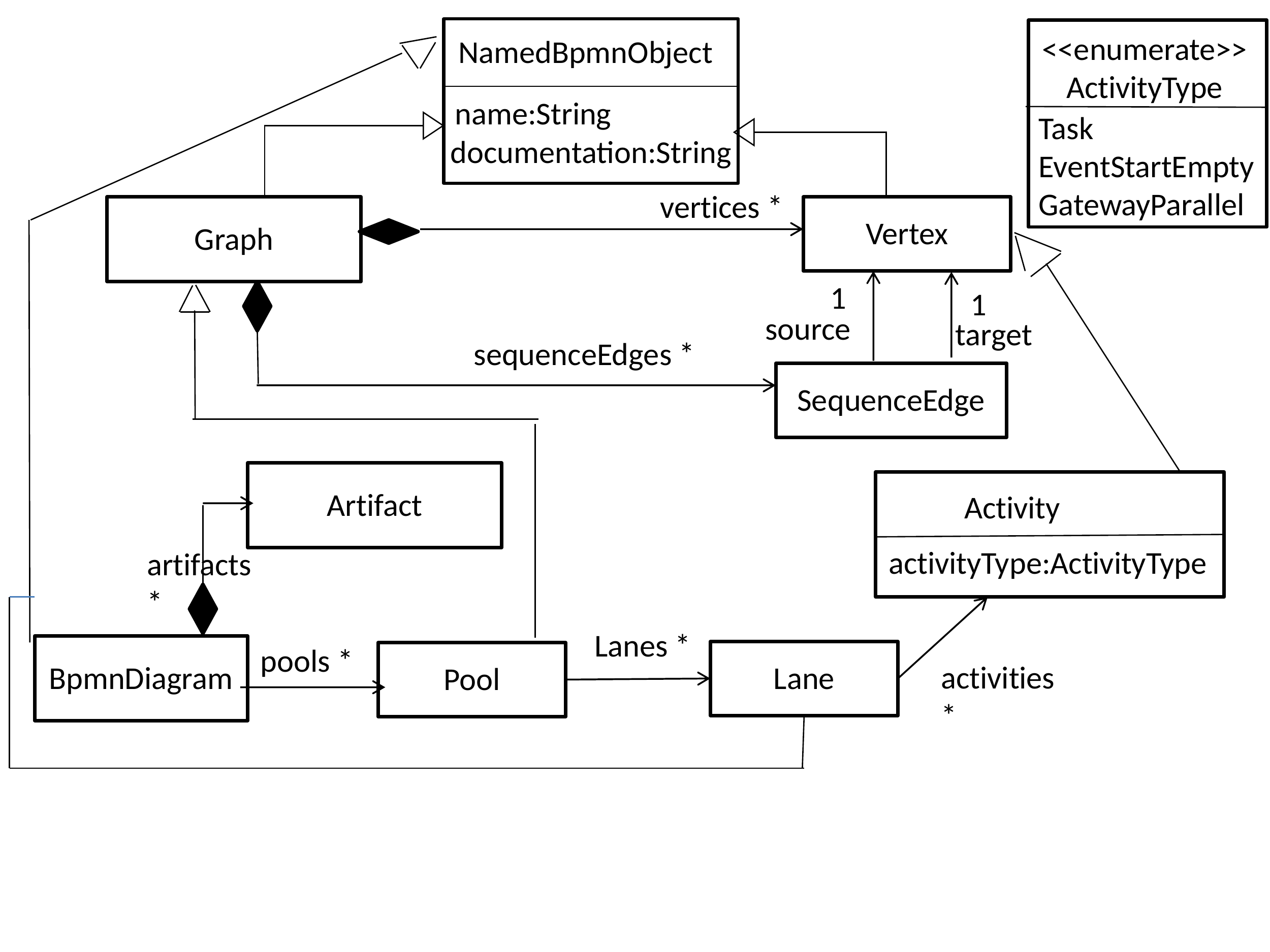}
        \caption{Intalio BPMN metamodel}
    \end{subfigure}
    \caption{Excerpts UML activity diagrams (AD) metamodel and Intalio Business process model (BPMN) metamodel}
    \label{fig:Motivatingexample}
\end{figure*}

Listing~\ref{lst:syntacticerrors} presents an excerpt of an ATL transformation program of UML activity diagrams into Intalio business process models\footnote{\url{http://www.intalio.com/products/bpms}}, borrowed from~\cite{cuadrado2018-quickfix}. The two metamodels are shown in Fig.~\ref{fig:Motivatingexample}.

ATL transformation programs consist in a source metamodel (\texttt{IN}), a target metamodel (\texttt{OUT}), and a set of transformation \texttt{rule}s. 
Each rule is named and describes a pattern in the source metamodel (\texttt{from} part, also called the input pattern) and a pattern in the target metamodel (\texttt{to} part, also called the output pattern).
An ATL transformation program uses an execution mechanism triggering a rule when an object in the input model matches the input pattern  of the rule.
When the rule is executed, an object is created in the output model according to the output pattern  of the rule.
For example, the rule \texttt{activity2diagram} (lines 7-12) states that each object instance of the \texttt{Activity} class of UML  (line 8) triggers the creation of an object, instance of the \texttt{BpmnDiagram} class of Intalio  (line 9). 

\begin{lstlisting}[
    breaklines=true,
    keepspaces=false,
    breakindent=0pt,
%    basicstyle=\ttfamily\footnotesize\scriptsize,
    basicstyle=\ttfamily\scriptsize,
    caption={Excerpt of an ATL transformation program, from UML Activity Diagram to Intalio BPMN},label={lst:syntacticerrors}]
1 create OUT : Intalio from IN : UML;
2 ...
3 helper context UML!Activity def: allPartitions      
4       :Sequence(UML!ActivityPartition) =  
5        self.partition->collect(p |  p.allPartitions)->flatten();
6
7 rule activity2diagram {
8   from    a : UML!Activity
9   to      d : Intalio!BpmnDiagram (
10              artifacts <- a.name,
11              pools <- a.allPartitions
12          )}
13
14  rule activitypartition2pool {		
15  from    a : UML!Comment
16  to      p : Intalio!Pool,	
17	        l : Intalio!Lane ( 	
18              activities <- a.node->reject(
19                  e|e.oclIsKindOf(UML!ObjectNode))
20          )}
21...
\end{lstlisting}

An input object may trigger the creation of several output objects.
For instance, the rule \texttt{activitypartition2pool} (lines 14-20) states that each object instance of the \texttt{Comment} class of UML (line 15) triggers the creation of two objects in the output model: one instance of the \texttt{Pool} class of Intalio (line 16) and the other instance of the \texttt{Lane} class of Intalio (line 17). 
Input and output objects are related by a trace link: it is possible to access properties of the input object and to set those of the output object.
For instance, the rule \texttt{activity2diagram} initializes the properties \textit{artifacts} and \textit{pools} of \texttt{BpmnDiagram} depending on properties it accesses in \texttt{Activity} (lines 10-11).
Property initialization, called \textit{binding} in ATL, may use a property of the input object (line 10), a \texttt{helper} (similar to methods, as defined in lines 3-5) to reshape the input object property (line 11), or  OCL constraints (lines 18-19).
Properties can be attributes with native types, or references towards objects.
When a binding's right-hand side (RHS) is a reference to an object of the input model, we cannot assign this object directly to the output object's property of the left-hand side (LHS).
In fact, the object of the input model needs to be transformed into elements of the output model.
In this case, a binding resolution mechanism is used to retrieve a rule which can perform this transformation, i.e., with a \texttt{from} part corresponding to the type of the input model object (binding's RHS), and a \texttt{to} part corresponding to the type of the output model property (binding's LHS).
Models are primary artifacts that are exploited through model transformations~\cite{sendall2003-MTHeartAndSoul}.
Transformations use a type system mostly defined by the source and target metamodels, \emph{i.e.}, the input and output pattern elements in transformations have to refer to existing elements in the involved metamodels~\cite{cuadrado2017-staticanalysis}.
Consequently, a type error can be introduced in a transformation program
by accident during development (developer or domain expert error) 
by wrongly using  the metamodel types.
It can also result from changes in the metamodels it uses, but this case is out of the scope of this paper.
Resolving type errors in ATL is thus difficult because of the declarative nature of the transformation language and the dependencies towards the involved metamodels. 
In the following, we illustrate type errors using the  transformation program excerpt of Listing~\ref{lst:syntacticerrors}.
A common type error concerns properties' types in bindings, such as in line 10.
In the Intelio metamodel, the property \textit{artifacts} refers to objects of type \texttt{Artifact}. 
However, in the RHS of this binding, the input object property \textit{name} is of type \texttt{String}, causing a \textit{incompatible type} error.
Invalid types are also frequent errors. As mentioned earlier, each rule is triggered by an input object that is compatible with the \texttt{(from)} part of the rule. 
The rule \texttt{activitypartition2pool} is thus triggered by objects conforming to \texttt{Comment} in the UML metamodel (line 15).
If we look closely to the UML metamodel of Fig.~\ref{fig:Motivatingexample}, there is no \texttt{Comment} element: this raises the error \textit{invalid type}. 
Another common error concerns the binding resolution.
Let us consider the binding of line 11. 
The RHS of the binding calls an helper returning objects of type \texttt{ActivityPartition} from UML metamodel.
The LHS of the binding is the property \textit{pool}, with type \texttt{Pool} from Intelio metamodel.
To resolve this binding, there must be a rule in the transformation program having \texttt{UML!ActivityPartition} as input pattern, and \texttt{Intelio!Pool} as output pattern, which is not the case in our excerpt: this raises an \textit{possible unresolved
binding}~\cite{cuadrado2018-quickfix} error.

\subsection{Challenges of Fixing Model Transformations}\label{sec:pbstatement}

Existing research on model transformation repair generally follows  the precept that errors sharing ``the same symptoms, the same root cause, or the same solution'' can be fixed in the same fashion~\cite{martinez2014-fixIngredientsRedundancyForRepair}. Concretely, for a range (or class) of equivalent errors, a predefined patch is applied to all instances of this kind of error.
Cuadrado et al. present an evolvable list of patches tailored as a response to every characterized type of syntactic error~\cite{cuadrado2018-quickfix}.
The authors point that the proposed list of patches may evolve with new error types or with the refinement of existing patches. 
Additionally, one may want to adapt patches to other transformation languages. 
Thus, defining, refining and adapting patches require an important amount of knowledge, thorough study of their impact and manual maintenance effort. 
Another issue of predefined patches mentioned by the authors is that the order in which one applies patches may bring unexpected interactions and side effects on the transformation, \textit{e.g} new errors can be injected or contradictory changes may loop. 

We do believe that an approach that explores dynamically candidate patches, rather than applying/instantiating predefined ones, can circumvent the above-mentioned issues.
As patches can be seen as sequences of basic edit operations, such an approach can automatically explore the space of possible sequences that fix several typing errors at once, without creating new ones. 
Another even more important issue arising when fixing type errors is to ensure that the original behavior/semantics of the transformation is not altered -- or at least that the semantic discrepancy is circumscribed and characterized. 
The common way to ensure this behavior preservation after changes is to use an oracle such as test suites, pre-/post-conditions, or possibly other specifications.
Yet, in the context of domain-specific  problems such as those MDE offers to solve, the necessary knowledge required to build a relevant and trustable oracle is not always available~\cite{baudry2010-barrierToTestingMT}. More important yet, since we are dealing with faulty (i.e., non executable) transformations, potentially good candidate patches cannot  be evaluated through test cases, since they may not correct all errors, resulting in corrected transformations that still cannot be executed.
In our approach, we propose to take additional behavior-preserving objectives into account, and thus view the exploration of the space of candidate patches as a multi-objective optimization problem.

\subsection{NSGA-II, a Multi-Objective Population-based Evolutionary Algorithm} \label{sec:NSGAII} 
\begin{figure}[ht]
	\centering
	\includegraphics[width=.9\linewidth]{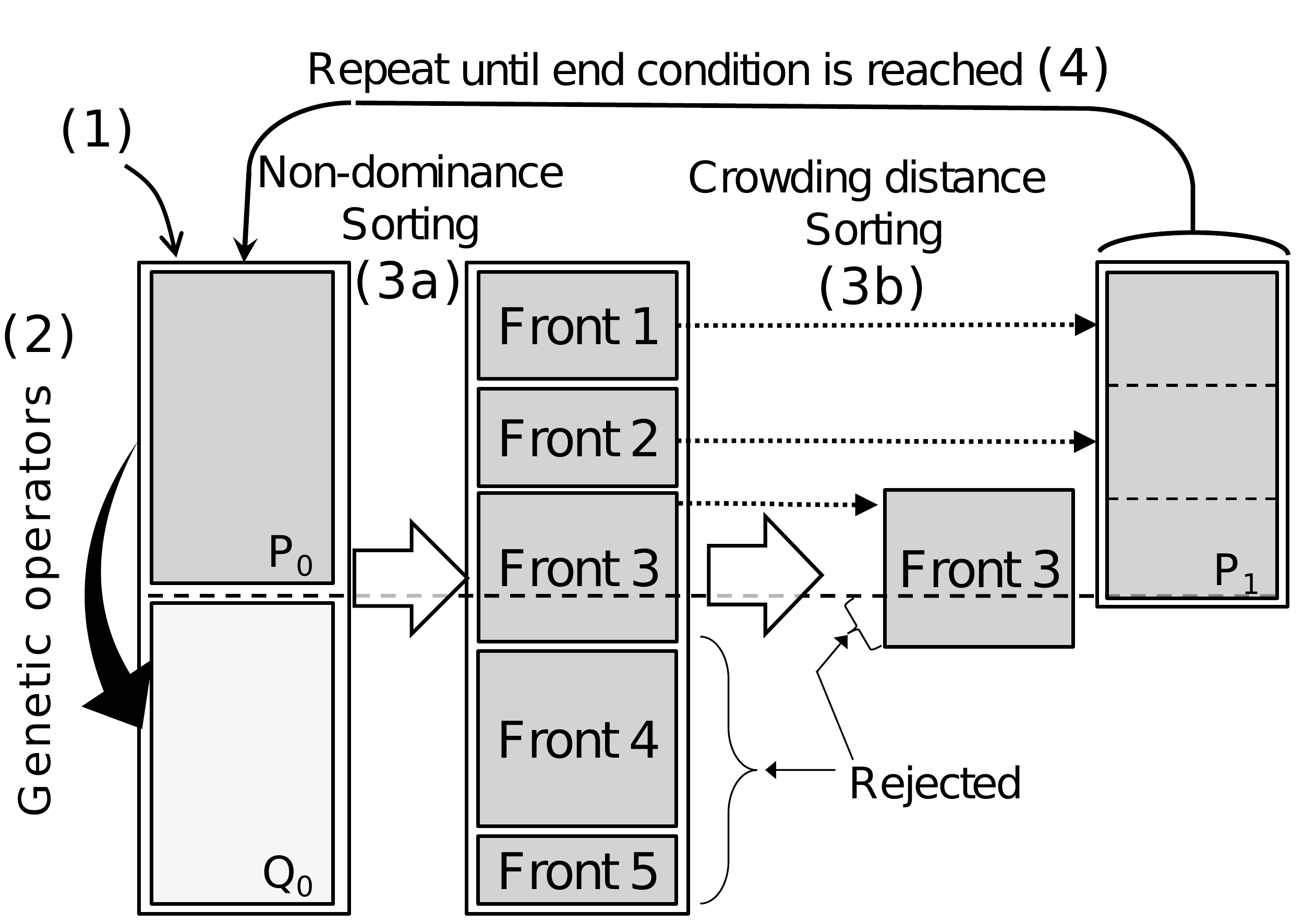}
	\caption{NSGA-II Algorithm~\cite{deb2000-nsga}}
	\label{fig:nsga2}
\end{figure}
For a multi-objective optimization problem, the idea of evolutionary population-based algorithms (EPA) is to make a population of candidate solutions evolve toward a near-optimal solution in order to solve that problem.
As such, EPA are designed to find a set of optimal solutions, called non-dominated solutions, or Pareto set.
A non-dominated solution provides a suitable compromise between all objectives such that one objective cannot be further improved without degrading another objective.

In this paper, we use NSGA-II, a well-known fast multi-objective genetic algorithm, that is suitable to the kind of problem we are solving~\cite{ali2020quality}.
The first step in NSGA-II as described in Fig.~\ref{fig:nsga2} is to create randomly a population $P_0$ of $N/2$ individuals encoded using a specific representation (1).
Then to complete the population of size $N$, a child population $Q_0$, also of size $N/2$, is generated from the population of parents $P_0$ using genetic operators such as crossover and mutation (2).
Both populations are merged into an initial population $R_0$ of size $N$, which is then sorted into dominance fronts according to the dominance principle (3a). 
A solution $s_1$ dominates a solution $s_2$ for a set of objectives $\{O_i\}$ if $\forall i, O_i(s_1) \geqslant O_i(s_2)$ and $\exists j \mid O_j(s_1) > O_j(s_2)$.
The first (Pareto) front includes the non-dominated near-optimal solutions.
The second front contains the solutions that are dominated only by the solutions of the first front, and so on and so forth.
The fronts are included in the parent population $P_1$ of the next generation following the dominance order until the size of $N/2$ is reached. If this size coincides with part of a front, the solutions inside this front are sorted, to complete the population, according to a crowding distance which favors diversity in the solutions~\cite{deb2000-nsga}  (3b). This process is repeated (4) until a stop criterion is reached, \textit{e.g.} a number of iterations or one or more objectives greater than a certain threshold. 
\section{Multi-step derivation of patches}
\label{sec:approach}
\subsection{Approach Overview}
\begin{figure}[htb!]
	\centering
	\includegraphics[width=.9\linewidth]{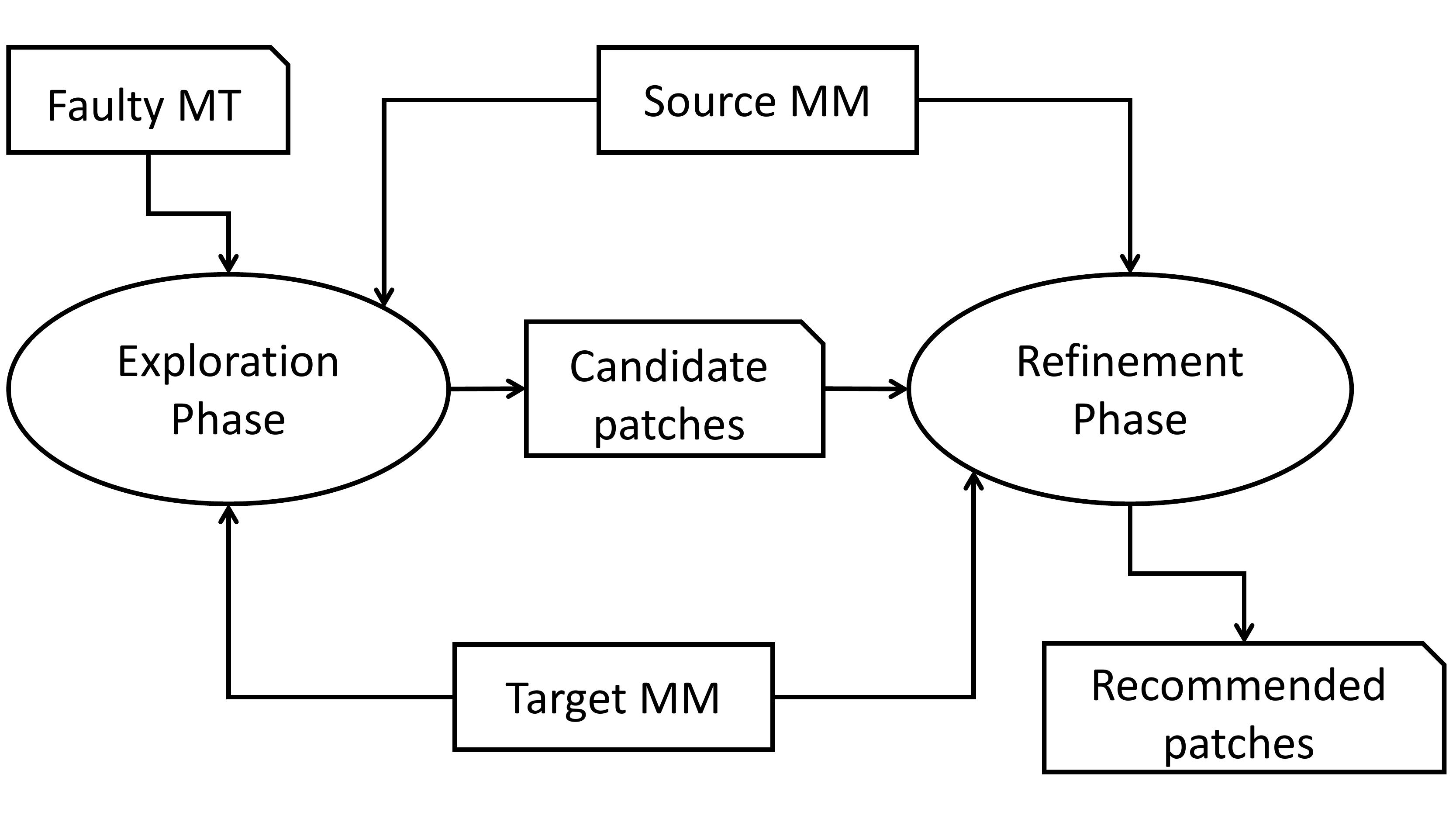}
	\caption{Overview of the proposed two-steps approach}
	\label{fig:approach_overview}
\end{figure}

We propose a two-steps approach to generate repair patches for transformations containing multiple type errors, as depicted in Fig.~\ref{fig:approach_overview}. The first step, ``Exploration phase'', takes as input a faulty transformation and the source and target metamodels defining the domain type system, and produces candidate patches. This step has two goals: (1) exploring the space of possible patches with the objective of correcting as much as possible type errors, and (2) minimizing the deviation from the original behavior by combining two lightweight surrogate objectives to testing.

As the first step is based on an evolutionary population-based algorithm, the exploration evaluates an important number of solutions. Consequently, we cannot afford to use resource-consuming objectives for behavior preservation. Alternatively, after a patch solution is produced by the first step, we refine it in a second step to increase the likelihood that type-error fixes do not alter the behavior. ``Refinement phase'' exploits four heuristics that better determine the parameters of some change operations included in the candidate patches or propose alternative change operations. In the remainder of this section, we describe both steps.      
\subsection{Exploration Phase}
\label{sec:ep}
In order to adapt NSGA-II, like any evolutionary population-based algorithm, to our problem, three key points must be defined: solution representation, solution derivation, and fitness evaluation.

\paragraph{Solution representation.}
As stated in the previous section, a patch can be seen as a sequence of edit operations that should be applied on the faulty transformation to correct it. 
We use  the basic edit operations listed in Table~\ref{t:oplist}~\cite{cuadrado2018-quickfix} to build our candidate sequences. 

\begin{table}[h]
\small
	\centering
	
	\caption{Set of basic edit operations of model transformations taken from ~\cite{cuadrado2018-quickfix}}
	\label{t:oplist}
	\begin{tabular}{ll}
		\textbf{Operator}                  & \textbf{Target}               \\ \hline
		Creation                  & Binding                                          \\
\hline

		Type          & Type of source/target pattern element            \\
		
		modification& Type of variable or collection                   \\
		& Type parameter (e.g., oclIsKindOf(Type))         \\ \hline
		Feature name  & Navigation expression    (binding RHS)                        \\
		modification & Target of binding     (binding LHS)                           \\ \hline
		Operation& Predefined operation call (e.g., size)                \\
		modification& Collection operation call (e.g., includes)            \\
		& Iterator call (e.g., exists, collect)                 \\	\hline	\color{gray} Deletion   & 	\color{gray} Rule, helper, binding  ... 
	    \\ 
	&
	\end{tabular}
\end{table}

The operation \textit{binding creation} adds a new binding in a rule.  
\textit{Type of source/target pattern element} changes the type of the \texttt{from} or \texttt{to} part of a rule.
\textit{Type of variable collection} changes the type of a collection such as the type \texttt{UML!ActivityPartition} of the Sequence in line 4 of the listing.
The operation \textit{Type parameter} changes the parameter \textit{Type} of a function such as \texttt{oclIsKindOf()}.
\textit{Navigation expression} and \textit{Target of binding} replace, respectively, the RHS and LHS of a given binding.
\textit{Predefined operation call modification},  \textit{Collection operation call modification} and
\textit{Iterator call modification} replaces a function call by another one (e.g., \texttt{collect()} or \texttt{flatten()} from line 5).

As we are dealing with type errors in transformations, it is important to pay a special attention to the delete operators. Indeed, sequences with these operators may artificially resolve some errors by removing faulty fragments of statements, statements or rules containing errors.
Therefore, we ignore delete operators at this stage of our work. 
For the sake of consistency, in our evaluation in
Section~\ref{sec:evaluation}, we do not consider errors that
require  delete operations.

\begin{figure*}[ht]
	\centering
	 \includegraphics[width=.8\linewidth]{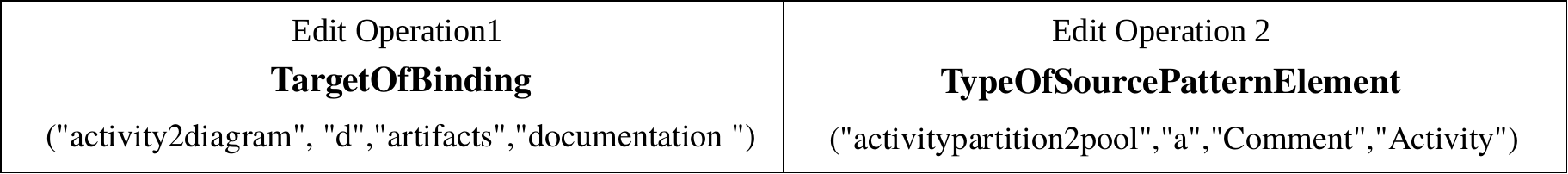}
	\caption{Example of a sequence of two edit operations (patch) which can be applied on the ATL transformation program of Listing~\ref{lst:syntacticerrors}}
	\label{fig:solrepresentation}
\end{figure*}

Fig.~\ref{fig:solrepresentation} presents a example of a sequence with two edit operations (i.e., a patch) which can be applied on Listing~\ref{lst:syntacticerrors} to fix some of the type errors identified in Section~\ref{sec:context}. 

In candidate sequences, each edit operation is identified by a name as defined in Table~\ref{t:oplist}.
These two operations have four parameters: \textit{ruletoModify}, \textit{objectToModify}, \textit{oldValue}, and \textit{newValue}.
For example, the edit operation \texttt{TargetOfBinding} changes the target of the binding (its LHS) from \textit{artifacts} to \textit{documentation} in the rule \texttt{activity2diagram} in line 10. 
The edit operation \texttt{TypeOfSourcePatternElement} replaces \textit{Comment} by \textit{Activity} in the input pattern of the rule \texttt{activitypartition2pool}.
Applying this patch on Listing~\ref{lst:syntacticerrors} produces the Listing~\ref{lst:semanticerrors} in which two type errors of different categories have been simultaneously corrected: the \textit{incompatible type} error of line 10 is fixed as properties \textit{documentation} and \textit{name} are both of type \texttt{String}, and the \textit{invalid type} error of line 15 is fixed as \texttt{Activity} is an existing element of the source metamodel. 

A solution is then defined by selecting a sequence of operations and by assigning values to their parameters. The solution space thus spans over all potential combinations of operations, their parameterizations and their order.

\begin{lstlisting}[
    breaklines=true,
    keepspaces=false,
    breakindent=0pt,
%    basicstyle=\ttfamily\footnotesize\scriptsize,
    basicstyle=\ttfamily\scriptsize,
    caption={Model transformation program of Listing~\ref{lst:syntacticerrors} after applying the patch of Fig.~\ref{fig:solrepresentation}},label={lst:semanticerrors}]
1 create OUT : Intalio from IN : UML;
2 ...
3 helper context UML!Activity def: allPartitions      
4       :Sequence(UML!ActivityPartition) =  
5        self.partition->collect(p |  p.allPartitions)->flatten();
6
7 rule activity2diagram {
8   from    a : UML!Activity
9   to      d : Intalio!BpmnDiagram (
10              documentation <- a.name,
11              pools <- a.allPartitions
12          )}
13
14 rule activitypartition2pool {		
15  from    a : UML!Activity
16  to      p : Intalio!Pool,	
17	        l : Intalio!Lane ( 	
18              activities <- a.node->reject(
19                  e|e.oclIsKindOf(UML!ObjectNode))
20          )}
21 ...
\end{lstlisting}

\paragraph{Solution derivation.} 

Two kinds of operators derive new solutions from existing ones: \textit{crossover}, i.e., the recombination of the existing genetic material, and \textit{mutation}, i.e., the injection of new genetic material. 
A sequence of operations is a convenient representation for breeding through genetic operators. 
In our adaptation, we use single point crossover operator. This operator consists in cutting the operation sequences of two selected solutions into two parts and in swapping the parts at the right of the cut point to create two new solutions.
The mutation operator introduces random changes into candidate solutions. 
In our adaptation, it selects one or more operation(s) from the solution sequence and either replaces them by another type of edit operation or modifies the parameters. 

\paragraph{Fitness evaluation.}  
A good solution is a sequence of operations which, when applied on a transformation, \textit{(\textit{i}) fixes the type errors} and \textit{(ii) preserves the transformation behavior}. 
The objective of fixing type errors can be directly evaluated by tools based on transformation language features, such as static fault analysis. This is our first objective:

\textbf{(1) Fixing type errors} \textit{or, to minimize the number of transformation errors.} We used this objective to check the number of errors in the transformation rules after applying the sequence of change operations. To measure the number of errors, we use the AnATLyzer tool~\cite{cuadrado2018-anatlyser}, which finds a wide range of syntactic errors (including type errors) in ATL transformations using static analysis.
	Formally, the objective function for a solution S is:
	$Minf1(S)=|Errors(S)|$.
	
The behavior-preservation objective poses a significant challenge and is difficult to capture with a single objective. 
In this paper, we explore the combination of two additional objectives that we believe favors behavior-preservation:
 \textbf{(2) To favor solutions of small size}  \textit{or, to minimize the number of change operations.}  This objective represents the number of operations in (or the size of) a candidate patch sequence. We used this objective to reduce the deviation from the initial transformation, and then the risk of changing the semantic. Additionally, we want to prevent the solutions to grow unnecessary large and escape the bloating effect~\cite{dejong2002-bloating}. Formally:
	$Minf2(S)=|V(S)|$,
	where V is the solution's sequence of operations.	
	
\textbf{(3) To keep changes local} \textit{or, to minimize the alteration of the metamodels' footprint:} The footprint of the source or target metamodels defined by an (initial or candidate) transformation is estimated by the number of elements from both source and target metamodels that the candidate solution employs (resp. does not employ) whereas the original transformation does not (resp. employs)~\cite{burgueno2015-staticfaultlocalization}.
	Formally, the third objective can be expressed as follows:
	$Minf3(S)=|SFP(O)-SFP(S)| + |TFP(O)-TFP(S)|$,
	where SFP and TFP are the footprints in respectively the source and target metamodels, extracted from the original transformation O and the candidate transformation S. To extract the footprint set of a transformation for a metamodel, we use the footprint tool defined by Burgueño et al.~\cite{burgueno2015-staticfaultlocalization}.

These objectives are conflicting in essence. This is why we solve the multi-objective patch derivation problem by adapting the  evolutionary population-based algorithm NSGA-II~\cite{deb2000-nsga}  described in Section~\ref{sec:NSGAII}.

\subsection{Refinement Phase}

The exploration phase produces a set of candidate patch solutions corresponding to the Pareto front of the last iteration of NSGA-II. These solutions may remove completely or partially syntactic type errors detected by AnATLyzer, but do not semantically correct some of these errors. There are many reasons that could explain this phenomenon. For example, the choice of a parameter for a given change operation is made without checking the global consistency with the other change operations in the sequence. Another example is when many type-compatible possibilities exist for a given parameter, one is selected randomly without a proper way to evaluate the likelihood of each possibility to semantically correct the error. 

One can sophisticate the decision process of operation and parameter choices in the exploration phase, but this comes at high computation cost considering the number of explored solutions. Thus, we decided to alternatively refine one or more solutions produced by the exploration phase. As the refinement concerns a few solutions, we can afford a more resource-consuming decision process. 

After analyzing the used change operations, we identified four for which we can define heuristics to improve the decisions made during the exploration phase. 
In what follows, we present the improvement heuristics for these operations and illustrate their mechanisms on transformation program excerpts rather than sequences of edit operations, as it is easier to comprehend.

\begin{figure}[ht]
    \centering
    \includegraphics[width=.8\linewidth]{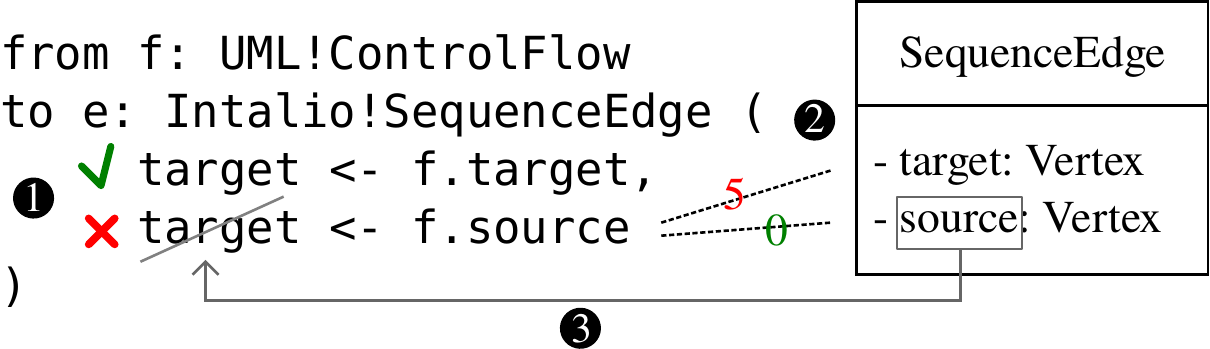}
    \caption{Heuristic for the operation TargetOfBinding}
    \label{fig:targetOfBinding}
\end{figure}

\textbf{(1) Target of binding.}
This edit operation changes the LHS of a binding. It may thus produce several bindings having the same target property in a given rule, even though a property should not be initialized more than once.
In these cases, the heuristic seeks to change the LHS of necessary bindings until no property is initialized several times, as illustrated in Fig.~\ref{fig:targetOfBinding}. 
First, the heuristic performs an edit distance computation between the target property and the different RHS properties: the binding with the minimum distance is ignored on the next step as it is considered the correct initialization (\ding{182}).
Then, the heuristic retrieves the list of accessible target properties, and computes the edit distance with each RHS of the remaining bindings (\ding{183}).
Finally, it modifies the target properties with the closest property of this list (\ding{184}).

When the binding RHS is of type \texttt{String}, changing the LHS to any \texttt{String} property prevents  typing errors.
However, it is common to select an incorrect LHS with regards to semantics.
Steps \ding{183} and \ding{184} can be applied in this particular case.
Applying this heuristic in Listing~\ref{lst:semanticerrors} would change the binding 
\texttt{documentation <- a.name}  (line 10) to 
\texttt{name <- a.name}, which is more coherent.

\begin{figure}[ht]
    \centering
    \includegraphics[width=.8\linewidth]{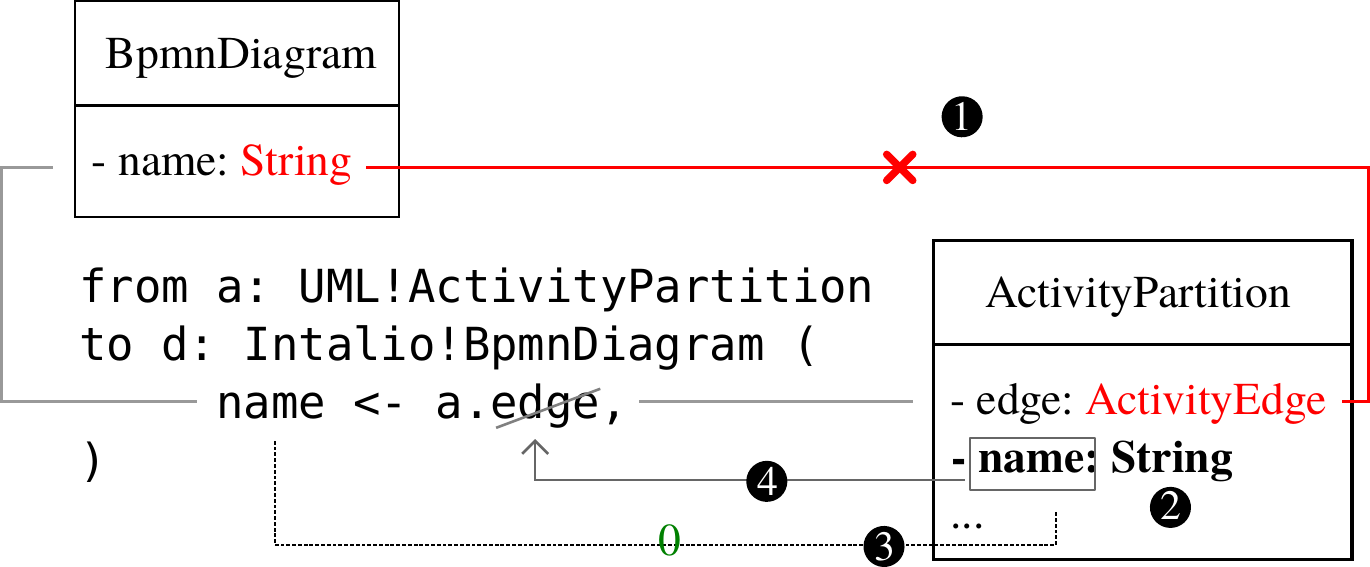}
    \caption{Heuristic for the operation navigationExpression}
    \label{fig:navigationExpression}
\end{figure}

\textbf{(2) Navigation expression.} This operation may change a binding's RHS to have a different type than the binding's LHS (see Fig.~\ref{fig:navigationExpression}), causing a \textit{type mismatch} error (\ding{182}).
When the property in the binding's LHS is of type \texttt{String}, the heuristic first retrieves the list of accessible properties which are of \texttt{String} type in the input model element (\ding{183}). Then, it selects from this list the property's name having the smallest edit distance with the LHS property's name (\ding{184}) and replaces the RHS accordingly (\ding{185}).

\textbf{(3) Type of source (target) pattern element.}
This edit operation may introduce an improper type in the \texttt{from} part of a rule. 
We have seen previously that each binding which takes into account references towards objects should be resolved (i.e., associated with the correct rule in the transformation). 
The correct rule has its \texttt{from} part corresponding to the type of the RHS of the binding and it \texttt{to} part corresponding to the type of the LHS of the binding.
We could use the RHS of the binding that refers to that rule to infer the correct type for the \texttt{from} part.
The third heuristic checks existing bindings to find the one with a LHS which type is equivalent to the \texttt{to} part of a given rule (\ding{182}).
Then, it verifies if the type of the RHS of this binding corresponds to the \texttt{from} part of the rule, and changes the latter accordingly if it is not the case (\ding{183}).
This heuristic can be applied when there exists only one rule having a given type in its \texttt{to} part.

\begin{figure}[ht]
    \centering
    \includegraphics[width=.9\linewidth]{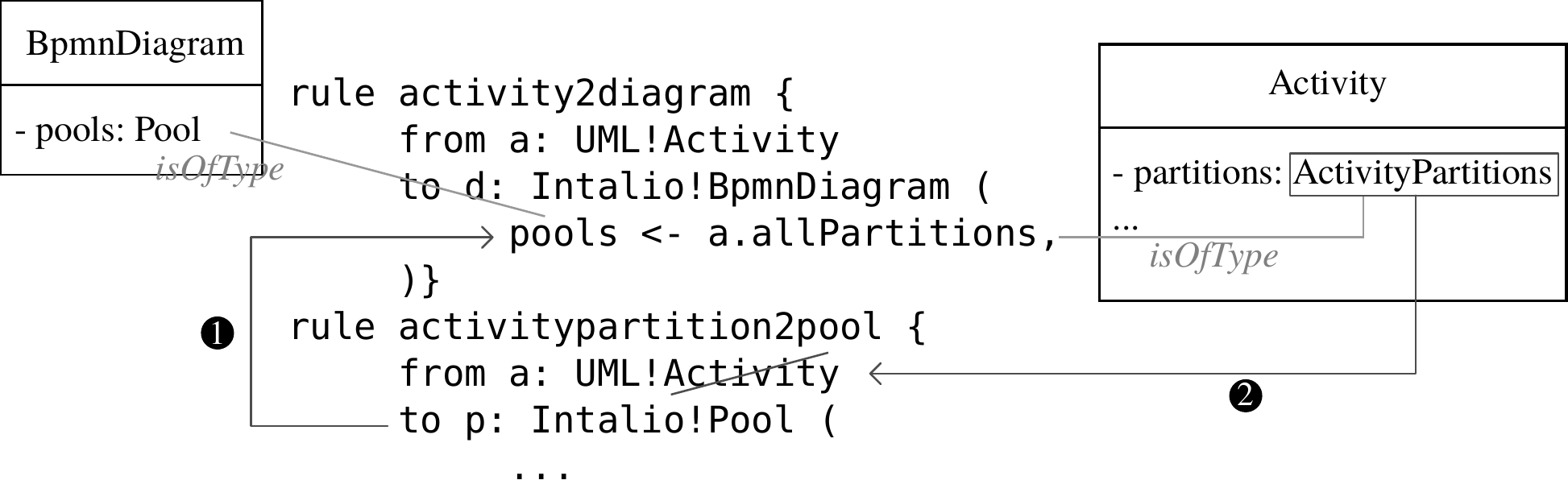}
    \caption{Heuristic for the operation typeOfSourcePatternElement}
    \label{fig:sourcePatternElement}
\end{figure}

\begin{figure}[ht]
	\centering
	 \includegraphics[width=1\linewidth]{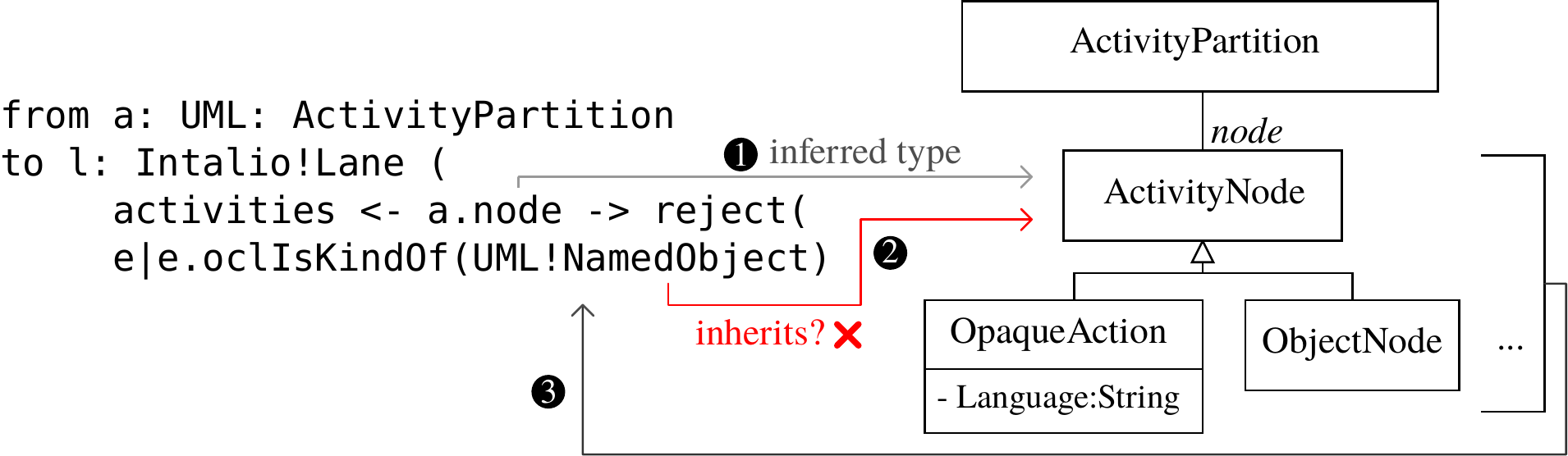}
	\caption{Heuristic for operation TypeParameter (e.g., oclIsKindOf(Type))}
	\label{fig:argument-heuristic}
\end{figure}

The same verification can be done the other way around for verifying the \texttt{to} part of the rule. 
\textbf{(4) Type parameter (e.g., oclIsKindOf(Type)).}
This edit operation change the \texttt{Type} parameter defined in functions such as \texttt{oclIsKindOf} or \texttt{oclAsType}. 
Based on OCL definition, \texttt{Type} parameter
 of \texttt{oclIsKindOf}, for example, must inherit from the type  defined before (i.e., the inferred type).
 For instance, in Fig.~\ref{fig:argument-heuristic}, \texttt{UML!NamedObject} should inherit the inferred type of \texttt{a.node}, i.e., \texttt{ActivityNode}.

The fourth heuristic first retrieves the inferred type (\ding{182}), then checks whether the \texttt{Type} parameter inherits from this type.
If not (\ding{183}), the heuristic changes the \texttt{Type} parameter by a subclass of the inferred type (\ding{184}). 
If \texttt{oclIsKindOf} is followed by  a property, e.g., \texttt{(e|e.oclIsKindOf(UML!NamedObject)) -> select(e|e.Language)}, the heuristic chooses a \texttt{Type} which has access to that property (here \texttt{OpaqueAction}). \\

Listing~\ref{lst:semanticerrors} contains two semantic errors (on lines 10 and 15) that are fixed by heuristics 1 and 3. This produces the transformation in Listing~\ref{lst:noerror}, now free of type error and behavior deviation.

\begin{lstlisting}[
    breaklines=true,
    keepspaces=false,
    breakindent=0pt,
%    basicstyle=\ttfamily\footnotesize\scriptsize,
    basicstyle=\ttfamily\scriptsize,
    caption={Model transformation program of Listing~\ref{lst:semanticerrors} after applying the heurisitics},label={lst:noerror}]
1 create OUT : Intalio from IN : UML;
2 ...
3 helper context UML!Activity def: allPartitions      
4       :Sequence(UML!ActivityPartition) =  
5        self.partition->collect(p |  p.allPartitions)->flatten();
6
7 rule activity2diagram {
8   from    a : UML!Activity
9   to      d : Intalio!BpmnDiagram (
10              name <- a.name,
11              pools <- a.allPartitions
12          )}
13
14 rule activitypartition2pool {		
15  from    a : UML!ActivityPartition
16  to      p : Intalio!Pool,	
17	        l : Intalio!Lane ( 	
18              activities <- a.node->reject(
19                  e|e.oclIsKindOf(UML!ObjectNode))
20          )}
21 ...
\end{lstlisting}
\section{Automatix - Preliminary Tool and Evaluation}\label{sec:evaluation}

 We implemented our approach in a tool, called Automatix, and performed an empirical
evaluation\footnote{For the review process, the experimental data can be downloaded using the link \url{https://bitbucket.org/zahravaraminy/ecmfa2021/src/master}}. The rest of this section describes the investigated research questions, details the evaluation procedure used, presents the results, and discusses the threats to the validity of our evaluation. 

\subsection{Reseach Questions}

As we explore many solutions during our evolutionary algorithm, it is legitimate to question whether the results are due to our search strategy or to the amount of candidate solutions explored during the search.
Thus, we start by performing a sanity check to compare the number of type errors fixed by patches obtained with our approach during the exploration phase and by patches obtained with a random search.
Then, we assess whether the patches obtained after the exploration phase preserve the behavior of the transformations.
Note that we do not evaluate the behavior of the output models that can be generated with a corrected transformation, but the behavior of the transformation program itself.
Finally, we do the same evaluation, but for patches obtained after the refinement phase.
In summary, we formulate the following research questions:

	\textbf{RQ0}: Are our results attributable to an efficient exploration of the search space,
or are they due to the large number of candidate solutions we explore during the evolution? 

	\textbf{RQ1}: Is the exploration phase able to correct type errors in transformations  while preserving their behavior?

\textbf{RQ2}: Is the refinement phase (combined with the exploration phase) able to correct type errors in transformations  while preserving their behavior?

\subsection{Evaluation Setup}

\begin{figure*}[htb!]
	\centering
	\includegraphics[width=.7\linewidth]{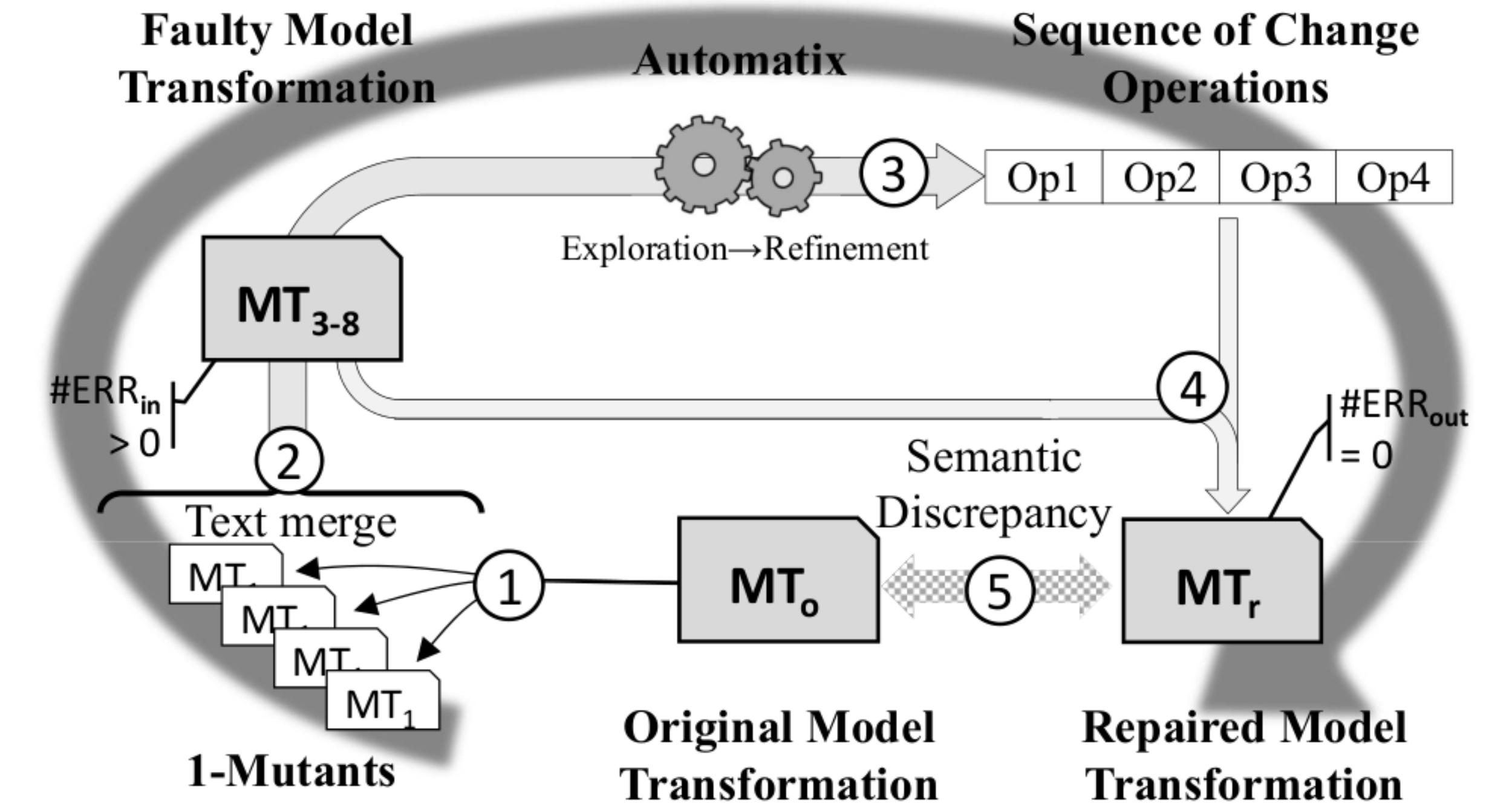}
	\caption{Evaluation procedure overview}
	\label{fig:evalbp}
\end{figure*}

To assess our approach's performance, we followed a rigorous protocol depicted in~Fig.~\ref{fig:evalbp}.

Steps 1 and 2 correspond to the creation process of faulty transformations from existing correct model transformations.
Then, step 3 represents the patch generation with Automatix, and step 4 the application of these patches on the faulty transformations to correct them.
Finally, step 5 is a manual comparison of the repaired model transformations against the original correct model transformations, which will help answer RQ1 and RQ2.
The sanity check (RQ0) uses the output of step 2 to perform a random search, and compare the results with the output of step 4.

We applied our evaluation on three existing third-party transformations, \textit{Class2Table}, \textit{PNML2PN} and \textit{UML2BPMN}
from the ATL Zoo\footnote{\url{http://www.eclipse.org/atl/atlTranformations}}.
\textit{Class2Table} takes as source a class diagram and outputs a relational database schema. \textit{PNML2PN} enables to produce a Petri net from an XML Petri net representation in the PNML format.
Finally, \textit{UML2BPMN} transforms a UML Activity Diagram into a business process model (Intalio BPMN) \cite{schumbacher2013-grafcet}.

To limit introducing bias in our evaluation, we used a list of existing faulty transformation mutants provided by the QuickFix
project~\cite{cuadrado2018-quickfix} (Fig.~\ref{fig:evalbp}, step 1).
Each mutant $MT_{1}$ corresponds to the original correct transformation $MT_{o}$ in which one error of a given class was injected, among the type error categories in ATL
transformations~\cite{cuadrado2018-quickfix}.

To create transformations with multiple errors, we selected randomly, for each of the three
transformation problems, 6 sets with
respectively 3 to 8 mutants coming from distinct error categories. 
Then, we merged
the mutants in each set to form 6 faulty transformations $MT_{3-8}$ with various numbers of type errors 
(Fig.~\ref{fig:evalbp}, step 2).  
Note that we performed the merge sequentially and,
then, the number of errors in the resulting transformations can be lower or higher
than the number of merged mutants, 
as some errors may overlap or create new errors as side effects.

This allowed us to consider faulty transformations with different numbers of errors, from all error categories, except those requiring delete operations to be fixed, as explained in  Section~\ref{sec:ep}.

Additionally, as we are using a probabilistic approach, we run our approach 5 times for each faulty transformation. Thus, we obtained for each transformation problem
30 different runs (5 runs * 6 faulty transformations).
For each faulty transformation, we created an initial population of 50 solutions,
\emph{i.e.,} sequences of edit operations, generated randomly. The edit operations
are generated for rules with flagged type errors. We complete the initial population
by 50 additional solutions obtained by crossover and mutation. We limited the number
of generations to 500, which means that, for each run, our algorithm explores
50,000 possible solutions (500x100). 
For the other parameters, the crossover and mutation rates
are respectively set to 0.8 and 0.2, values that usually perform well~\cite{haupt2004practical}.

In this evaluation, Automatix takes as input a faulty transformation, 
and produces candidate patches in the form of sequences
of change operations in two phases: one of exploration, and another of refinement (Fig.~\ref{fig:evalbp}, step (3)).
 To perform the evaluation, we define the following independent variables w.r.t. to the faulty transformations:
\begin{itemize}
	\item \textbf{\#MUT} -- Number of mutations applied on the original transformation to derive the faulty one.
	\item \textbf{\#ERR$_{in}$} -- Number of type errors found on the transformation after \#MUT mutations were applied.
\end{itemize}
We then define dependent variables w.r.t. the obtained candidate patches: 
\begin{itemize}
	\item \textbf{\#ERR$_{out}$} -- Number of type errors found on the transformation after a recommended patch has been applied.

	\item \textbf{\#OPE} -- Size of a recommended patch in number of change operations.
	\item \textbf{\#ITE} -- Number of algorithm iterations  before a recommended patch is found, \textit{i.e.} $\#ERR_{out(patches)} = 0$.
	\item \textbf{SEM} -- Rate of errors corrected while preserving the behavior, after the exploration phase (\textbf{EP}) and after the refinement phase (\textbf{RP}).

\end{itemize}

We used AnATLyzer to detect the type errors in the input and output transformation programs. This tool also allows us to identify which type errors of the input faulty transformation have been corrected.

Although, the exploration phase may produce more than one solution in the Pareto set, we decided to select only one for the refinement and for the comparison with the random search. To this end, we first select the solution that fixes the highest number of type errors according to AnATLyzer. In the case of a tie, we choose one with the shortest change-operation sequence. The two criteria were enough to reduce the possibilities to only one solution for all the runs.  

\subsection{Evaluation Results}

Table~\ref{rq1:table} summarizes the results of the different runs of our approach on the mutant
configurations described in the setup, for respectively \textit{Class2Table} , \textit{PNML2PN} and \textit{UML2BPMN}
transformations (col. 1). Except for \textbf{\#MUT}, the values indicate the average for the 5 runs on each faulty transformation. For \textbf{\#ERR$_{out}$}, we give the min, average and max values for the 5 runs.
Results are ranked by the number of mutations (\#MUT).
On average, the majority of errors introduced by the mutants (\#ERR$_{in}$ - col. 3), were successfully corrected
(\#ERR$_{out}$ - col. 6-8 indicating the min, average and max of the number of errors left after applying the patches found by Automatix), according to AnATLyzer.
We checked manually that the errors left are those intially introduced and not newly created ones by the patches.
For all the cases, we obtained at least one solution without any error left ($min = 0$).
Additionally, we did not observe a significant correlation between the number of inputs errors/mutants and the number
of generations to find a solution (\#ITE - col. 4).

\begin{table*}[h!]
	\centering
	\caption{Results of Automatix. Values are averages, unless precised.}
	\begin{tabular}{@{}c|ll|ll|ccc|l|l@{}}
		                       & {\#MUT} & {\#ERR$_{in}$} 	& {\#ITE}     	& {\#OPE} & \multicolumn{3}{c|}{\#ERR$_{out}$}    & \multicolumn{2}{c}{SEM}\\
		                              &      &      			&          	&       & min. & avg. & max.                  	 & EP &  RP  \\ \hline
		                              & 3    & 3,4              & 134       & 3     &  0   	& 0   & 0   			 & 68\% &68\%                         \\
		                              & 4 	 & 6   				&  44,8 	& 4 	&  0 	& 0   & 0   			 & 76\%   &\textbf{82\%}\\
		                              & 5 	 & 7,8 				&  56,8 	& 5	 	&  0 	& 0   & 0   			 & 91\%& \textbf{93\% }  \\
		                              & 6    & 8,4              & 226,8		& 6 	&  0  	& 0   & 0    			 & 91\%    &91\%                      \\
		                              & 7    & 9,6              & 261,8     & 5,8   &  0    & 0,6 & 1    					 & 71\%     & \textbf{75\%}                     \\
		{\multirow{-7}{*}{\rotatebox{90}{Class2Table }}}
									  & 8    & 10,6             & 276,5     & 7     &  0    & 0,2 & 1    					 & 87\% & 87\%                          \\ \midrule
		
		                              &  3   & 5,8  			& 86,4     	& 3,2   &  0   & 0    & 0              & 78\% &\textbf{89\% }                        \\
		                              &  4   & 7   				& 137,2 	& 4,2 	&  0   & 0,4  & 2    					 & 72\%  &\textbf{79\% }\\
		                              &  5   & 8,4 				& 188   	& 5,2 	&  0   & 0    & 0    			 & 80\%& \textbf{86\% }  \\
		                              &  6   & 9,2 				& 179,2     & 5,8   &  0   & 0,2  & 1    					 & 76\%  &\textbf{87\% }                        \\
		                              &  7   & 8,4              & 78,6      & 6     &  0   & 0,8  & 1   					 & 69\%     &\textbf{76\% }                    \\
		{\multirow{-7}{*}{\rotatebox{90}{PNML2PN    }}}
									  & 8    & 9,8              & 244,6     & 7,4   &  0   & 0,4  & 1    					 & 67\%       &\textbf{78\%}                   \\ \midrule

		                              & 3    & 3,2              & 162,4       & 3     &  0   	& 0   & 0   			 & 45\%      & \textbf{83\% }                   \\
		                              & 4 	 & 4,2   				&  35 	& 3.4 	&  0 	& 0,6   & {1}    			 & 34\%  &\textbf{41\%} \\
		                              & 5 	 & 6,8 				&  116 	& 5,2	 	&  0 	& 0,2   & {1}    			 & 44\% &\textbf{72\% } \\
		                              & 6    & 6,8              & 19,2		& 5,4 	&  0  	& 0,8   & {1}    			 & 40\%             &\textbf{53\%}             \\
		                              & 7    & 7,8              & 229,6     & 5,6   &  0    & 0,8 & 2    					 & 25\%            &\textbf{52\%}             \\
		{\multirow{-7}{*}{\rotatebox{90}{UML2BPMN }}}
									  & 8    & 8             & 72     & 5,8     &  0    & 1,2 & 2    					 & 26\%      &\textbf{40\%}                   \\	
				
	\end{tabular}
	\label{rq1:table}

\end{table*}

\subsubsection{RQ0 - Sanity Check}
To perform the sanity check we limited ourselves to a sample of runs.
We considered faulty transformations with 2, 4, 6 and 8 mutants for the problem of \textit{Class2Table}.
We compare the results obtained by Automatix to those
of a random search for the considered transformations.
Since Automatix explores 50,000 solutions  for each run, the random exploration
also picks, for each run, the best individual from 50,000 solutions generated randomly
in the same way as for the initial population in Automatix. As for Automatix, the random
exploration was also performed 5 times for each faulty transformation.

\begin{table}[h]
	\centering

	\caption{Automatix vs random results for \textit{Class2Table}.}
	\label{rq0}
	\begin{tabular}{l|c|c|c|c|}
	
		\cline{2-5}
		& \multicolumn{2}{c|}{\begin{tabular}[c]{@{}c@{}}Average \#ERR$_{out}$ \\ Value \end{tabular}} & \multirow{2}{*}{\begin{tabular}[c]{@{}c@{}}Mann Witney\\ p-value\end{tabular}} & \multirow{2}{*}{\begin{tabular}[c]{@{}c@{}}Effect Size\\ Cohen's d\end{tabular}} \\ \cline{1-3}
		\multicolumn{1}{|r|}{\rotatebox{90}{\#MUT~} } & \emph{Automatix}  & \emph{RDN}      &   &  \\ \hline
		\multicolumn{1}{|r|}{2 } & 0.0 & 0.2 &  0.374          &   -   \\
		\multicolumn{1}{|r|}{4 } & 0.0 & 2.8 & \textless 0.001  &  10.6   \\
		\multicolumn{1}{|r|}{6 } & 0.4 & 5.8 & \textless 0.001  &  8.6    \\
		\multicolumn{1}{|r|}{8 } & 3.0 & 6.4 & \textless 0.001  &  3.24   \\ \hline
	\end{tabular}

\end{table}

As shown in Table~\ref{rq0}, the solutions obtained with Automatix correct on average
clearly more errors 
than random ones.
Except for transformations with two mutants (first line), the difference between the two
strategies is statistically significant (T-Test with a \textit{p-value} lesser than 0.001),
and with a high effect size (Cohen's d greater than 3)\footnote{According to Sawilowsky~\cite{sawilowsky2009new}, an effect size greater than 2 is considered as huge}.

\subsubsection{RQ1 - Error Correction after Exploration Phase (EP)}

We consider that an error is actually corrected  in a transformation program when the change brought by the patch matches the corresponding code fragment in the original correct transformation $MT_{o}$. 
To assess that (Fig.~\ref{fig:evalbp}, step 5), we followed a two-steps process.
We started by applying an automated text diff between the original transformation $MT_{o}$ (the ground truth) and the transformation $MT_r$ fixed by a patch obtained after the exploration phase.
Then, we manually checked the discrepancies flagged by the diff to determine the number of errors that were corrected
without altering the behavior of the transformation (call them semantically fixed) and reported the rate of these errors  with respect to \#ERR$_{in}$ in columns SEM(EP). 
In this way, we are sure to determine if the applied patches obtained automatically are correcting type errors actually and not just making AnATLyzer not detecting them.

As shown in column SEM(EP) of Table~\ref{rq1:table}, the actual correction rates were good for two transformation problems. Indeed, we succeeded to correct on average between 68\% and 91\% of errors for \textit{Class2Table},
and between 67\% and 80\% for \textit{PNML2PN}.
As expected, some classes of type errors 
are difficult to correct syntactically while keeping a correct semantic, 
such as \textit{Invalid type}
and \textit{Compulsory feature not found}. These classes of errors require substituting or adding one of the many
features present in the metamodels with compatible types. This increases obviously the risk of choosing a wrong feature.

For the third transformation problem \textit{UML2BPMN}, the results were less good with an average actual correction rate
between 25\% and 45\%, although some executions reached higher scores. When analyzing the semantic discrepancies,
we noticed that, in addition to the complexity of the involved metamodels, these make an intensive use of inheritance.
Fixing errors like \textit{Invalid type} and \textit{Compulsory feature not found} with   correct solutions is thus even more difficult in this case.

\subsubsection{RQ2 - Error Correction after Refinement Phase (RP)}

To answer question RQ2, we perform the same semantic discrepancy but this time on patches obtained after executing the refinement phase on the candidate patches generated by the exploration phase. 

As shown in column SEM(RP) of Table~\ref{rq1:table}, the results indicate an improvement of the correction rates in  all three transformation problems.
By comparing correction rates between the exploration phase SEM(EP) and refinement phase SEM(RP), we can see that the heuristics improve the correction with behavior preservation of the transformations (increased rates are shown in boldface in the table). 
For \textit{Class2Table}, the correction rate increased on average from 80.7\% to 82.7\%. Over the 6 faulty transformation, 3 witnessed a higher rate (transformation with 4, 5, and 7 mutants). The improvement was more important for \textit{PNML2PN}, for which the average correction rate jumped from 73.7\% to 82.5\%. In this case, all the faulty transformations  saw their correction improve.
Finally, for \textit{UML2BPMN}, we observed sizeable improvements of correction rates from 35.7\% on average to 56.8\%. Here again, the improvement concerned all the faulty transformations reaching a rate of 83\% for the transformation with 3 mutants.

In the rates shown, we do not include errors that were partially corrected thanks to the heuristics. For example, we observed that for some bindings, the RHS was actually corrected but not the LHS. This means that the impact of the refinement phase can be much higher than one indicated by the correction rates.  

In conclusion, we show that the proposed approach is able to correct multiple type errors at the same time. 
The evaluation reveals that the two behavior-oriented objectives of the first phase circumscribe the risk of behavior alteration. It also shows that the combination of exploration and refinement phases allows to generate patches that correct most of the type errors while preserving the behavior.
These results are evidence that deepening the analysis of edit operations and the possible behavior deviations they may introduce help guiding the search through new objectives or refinement heuristics.

\subsection{Threats to Validity}
There are some threats that may call into question the validity of our evaluation results.
First, the faulty transformations used in the evaluation contain mutants and not errors actually introduced by developers.
We used this external data set because it is independent from our project and was used to evaluate the state-of-the-art work.
Moreover, it covers  a large spectrum of error types. Finally, the random combination of basic mutants we used can be representative
of the randomness with which errors can be introduced by developers.

Another limitation of our work at this stage of our project, is that we do not consider some of the error types (mutations).
In addition to errors that require delete operations mentioned earlier in the paper, we do not handle errors on ATL transformation
helpers. We expect to extend our work in the near future to also consider both families of errors.

Our approach does not produce a single solution, but rather a Pareto set of solutions. For the sake of automated evaluation,
we selected from the Pareto set the solution with the minimum number of errors left. In the case of a tie, we choose the smallest solution
(\textit{i.e.} with minimum \#OPE). We did the same for the RQ0 with the random exploration. In a real setting, other solutions,
discarded for their larger size, can be presented to the user, as well, as alternative solutions. This can be done by using a diversity
strategy to propose a representative sample of solutions \cite{Batot17Heuristic}.

We performed a manual inspection of the solutions to evaluate the semantic discrepancy between fixed transformations and
 original ones. In future evaluations, we plan to use a test suite of pairs of input-output models to check to
which extent the solutions proposed by our approach handle correctly the test cases (assess the behavior preservation).
Of course, this is possible only for solutions with no type errors.

\section{Related Work}\label{sec:rw}

The work presented in this paper crosscuts two research areas: program repair in general and verification and validation of model transformations. In the following subsections, we discuss representative work of both areas.

\paragraph{Program Repair.} 
 There is a plethora of works that try to automatically fix bugs in programs using different approaches such as genetic programming~\cite{LeGoues2013}, machine learning~\cite{Jeffrey2009, Martinez2013} or SMT solvers~\cite{demarco:hal-2014}. Most of the existing work targets a specific type of errors such as buggy IF conditions, memory allocation errors and infinite loops~\cite{Goues2013, logozzo2012modular, Muntean2015, Perkins2009AutomaticallyPE, demarco:hal-2014}. To evaluate the patches, most of the approaches use test suites as oracle. However, other oracles such as specifications (pre-/post-conditions) were also explored~\cite{Pei2014}.
Although these approaches produced promising results, they cannot be used to fix transformation typing errors. As mentioned in Section
~\ref{sec:pbstatement}, test suites are difficult to use and specifications are often not available. Moreover, we aim at correcting simultaneously a variety of errors types.

\paragraph{Model Transformation verification and validation.}
In this research area, there are three families of work: transformation testing, verification and validation of transformations, and transformation repair.

In the first family, Gogolla et al. ~\cite{gogolla2011}, for example,  presented
a model transformation testing approach
based on the concept of Tract, which defines a set of constraints on
the source and target metamodels, a set of source-target constraints, and a tract test
suite, i.e., a collection of source models satisfying the source
constraints. Then, they
automatically generated input models and transformed the source model into the target model. Finally, they verified that the source/target models
satisfy those constraints. There are other approaches to test the model transformations using other techniques such as graph patterns ~\cite{Balogh2010}, model fragments ~\cite{mottu2008}, Triple Graph Grammars (TGGs) ~\cite{wieber2014} or a combination
of these approaches ~\cite{giner2009}.

In second family, for example, Troya et al.~\cite{troya2018-spectrumbased} presented the Spectrum-Based Fault Localization technique and used
the results of test cases to determine the probability of each rule of transformation to be faulty.  Similarly Burgueño et al.~\cite{burgueno2015-staticfaultlocalization} presented a static approach for detecting the faulty rules in model transformations. Their approach uses matching functions that automatically create the alignment between specifications and implementations. Oakes et al.~\cite{OakesTLW182018} presented one method 
to fully verify pre-/post-condition contracts on declarative portion of ATL model transformations. Their approach
transforms the declarative portion of ATL transformations into DSLTrans and uses a symbolic-execution to produce a set of path conditions, which represent all possible executions to the transformation. To verify the transformation,
they verify pre-/post-condition contracts on these path conditions.
Finally, Cuadrado et al.~\cite{cuadrado2014-uncovering}  presented a combining approach involving a static analysis and constraint solving to detect errors in
model transformations. They detected potentially problematic statements and then used a witness model to confirm
the erroneous statements. We also used this technique for calculating the number of errors, which is defined as a fitness function.\\

All the above-mentioned approaches allow to find behavior errors and/or localize the faulty rules/statements. However, they do not propose patches to repair the errors, which is the goal of our approach. Yet, they can be used upstream of our approach like we did with AnATLyzer.

For work that fixes transformation errors, we distinguish between errors generated by the evolution of metamodels as addressed by Kessentini et al.~\cite{kessentini2018-MM-MT-Coevolution} and errors introduced by the developers. For these errors, to the best of our knowledge, the only existing work is Quick fix~\cite{cuadrado2018-quickfix}, which allows the correction of detected errors in ATL model transformation.
In this approach, they used the static analyser presented in~\cite{cuadrado2014-uncovering} to identify errors in ATL model transformations. Then, they extended the analyser to generate a catalogue of
quick fixes, which depends on static analysis and constraint solving, for identified errors in ATL transformations.
Then, quick fixes propose changes in the transformation based on the kind of error. The user selects a suitable fix among the proposed ones and applies interactively.
The differences with our work are that we aim at fixing errors jointly without predefined patch patterns, and also we target to generate a patch without requiring
human assistance, except for accepting/rejecting the patches.

Model transformations are not the only MDE artefacts targeted by repair approaches. There are many research contributions to generate patches for various modeling artifacts. Models are those that gather much attention as evidenced by the study of Macedo et al. ~\cite{Macedo2017Feature}. Another example of MDE artifact repair is given by Hegedus et al.~\cite{hege2011} in which the authors used state-space exploration techniques to generate quick fixes for Domain-Specific Modelling Languages (DSMLs).

\section{Conclusion and future work}\label{sec:conclusion}
In this paper, we explored the idea of fixing type errors in model transformation programs without relying on predefined patches. 
Considering that a patch is a sequence of basic edit operations, our approach explores the space of candidate sequences guided by two families of objectives: correction of type errors and behavior preservation.
While the correction of type errors is relatively easy to measure using transformation language features, behavior preservation poses many challenges. 
To tackle these issues, we proposed a two-phase approach to find candidate sequences that limit behavior deviations.
The first phase combined two objectives during the exploration to approximate the behavior preservation: minimizing the size of the sequence and keeping the changes local.
During a second phase, we applied four heuristics on the obtained patches to improve
the decisions made during the exploration phase. 
An evaluation of our idea showed that the first phase corrected a majority of type errors for two transformation problems, \textit{Class2Table} and \textit{PNML2PN}, most of the time without altering the behavior. We also showed that refining the patches obtained after the exploration using the four proposed heuristics significantly improved the quality of the patches in terms of behavior preservation for the three transformation problems, including \textit{UML2BPMN}.
As a future work, we plan to further investigate alternative objectives for behavior preservation to achieve correct and complete patches.
We also envision to inject some heuristics when selecting edit operations (initial population generation and mutations) to decrease the probability of altering the behavior.
Finally, we aim at generalizing our approach to repair semantic errors.

\bibliography{bibliography}
\section*{About the authors}
\shortbio{Zahra VaraminyBahnemiry}{is a PhD student at  the department of Computer Science and Operations Research (GEODES Group) of the Université de Montréal (Canada) \editorcontactf[]{varaminz@iro.umontreal.ca}}
\shortbio{Jessie Galasso}{is a post-doc  at the department of Computer Science and Operations Research (GEODES Group) of the Université de Montréal (Canada). \editorcontactf[]{jessie.galasso-carbonnel@umontreal.ca}}
\shortbio{Houari Sahraoui}{is a professor at the department of Computer Science and Operations Research (GEODES Group) of the Université de Montréal (Canada).
\editorcontact[]{sahraouh@iro.umontreal.ca }}
\end{document}